\documentclass[sigconf]{acmart}

\usepackage{subcaption}
\usepackage{cleveref}

\usepackage{algorithm}
\usepackage{algpseudocode}
\usepackage{amsmath}
\AtBeginDocument{%
  }

\setcopyright{none} 
\copyrightyear{2025}
\acmYear{2025}
\acmDOI{XXXXXXX.XXXXXXX}

\acmConference[MICRO 2025]{The 58th IEEE/ACM International Symposium on Microarchitecture}{October 18--22, 2025}{Seoul, Korea}
\acmISBN{978-X-XXXX-XXXX-X/XX/XX}

\begin{document}

\title{Learnable Sparsification of Die-to-Die Communication via Spike-Based Encoding}

\author{\normalsize
  Joshua Nardone\textsuperscript{1},
  Ruijie Zhu\textsuperscript{1},
  Joseph Callenes\textsuperscript{2},
  Mohammed E. Elbtity\textsuperscript{3},
  Ramtin Zand\textsuperscript{3},
  Jason Eshraghian\textsuperscript{1} \\
  \textsuperscript{1}University of California Santa Cruz, Santa Cruz, CA, USA \\
  \textsuperscript{2}California Polytechnic State University, San Luis Obispo, CA, USA \\
  \textsuperscript{3}University of South Carolina, Columbia, SC, USA
}

\acmConference[]{}{}

\renewcommand{\shortauthors}{}

\begin{abstract}
Efficient communication is central to both biological and artificial intelligence (AI) systems. In biological brains, the challenge of long-range communication across regions is addressed through sparse, spike-based signaling, minimizing energy and latency. Conversely, modern AI workloads are  increasingly constrained by bandwidth, leading to bottlenecks that hamper scalability and efficiency. Inspired by the brain's ability to execute dynamic and complex local computations coupled with sparse inter-neuron communication, we propose heterogeneous neural networks that combine spiking neural networks (SNNs) and artificial neural networks (ANNs) at bandwidth-limited regions, such as chip boundaries, where spike-based communication reduces data transfer overhead. Within each chip, dense ANN computations maintain high throughput, accuracy, and robustness. While SNNs have struggled to algorithmically scale, our approach surmounts this long-standing challenge through algorithm-architecture co-design where learnable sparsity is employed for die-to-die communication by confining spiking layers to specific partitions. This composable design combines high ANN performance with low-bandwidth SNN efficiency. Evaluations on language processing and computer vision exhibit up to 5.3$\times$ energy efficiency gains and 15.2$\times$ latency reductions, surpassing both purely spiking and non-spiking models. As model size grows, improvements scale accordingly. By targeting the inter-chip communication bottleneck with biologically inspired methods, this approach presents a promising path to more efficient AI systems.  
\end{abstract}





\maketitle

\section{Introduction}
The rapid expansion of artificial intelligence (AI) workloads has led to increasingly large and complex neural networks deployed across distributed computing systems. This scaling has intensified the challenges associated with data movement, particularly across chip boundaries, where bandwidth limitations and energy consumption become significant bottlenecks~\cite{hennessy2019new}. As models grow, exemplified by state-of-the-art architectures like OpenAI's GPT-4 with over a trillion parameters~\cite{patel2023gpt4}, the inefficiency of dense, continuous data communication hinders scalability and performance.

Evolution has addressed the challenge of long-range communication in biological brains through sparse, spike-based signaling~\cite{laughlin2003communication}. Neurons transmit information only when necessary, minimizing energy consumption and latency. This efficient, event-driven communication enables complex cognitive functions within a limited power budget.

Current approaches to mitigating data movement challenges in AI workloads involve both architectural and algorithmic strategies. Architectural solutions include advanced memory technologies like 3D stacked memory~\cite{kurshan2024towards} and high-bandwidth memory (HBM)~\cite{lee2016high}, as well as improved interconnects and packaging techniques. However, they often face diminishing returns due to scaling constraints, such as thermal dissipation limits, fabrication complexity, and increased costs, making them less effective for addressing the exponential growth in data communication demands of large-scale AI systems.

Algorithmically, techniques such as model pruning and quantization remove connections or decrease their precision, which introduces a trade-off between efficiency and model accuracy~\cite{sze2017efficient}. Structured sparsity, such as mixture-of-experts (MoE), deactivate entire sub-networks to preserve model size while reducing model utilization. All techniques have proven highly useful, but they largely rely on static or procedural data-processing pipelines that treat all inputs uniformly. Consequently, they miss the potential gains from event-driven and data-dependent adaptivity.

Spiking Neural Networks (SNNs) offer a biologically inspired alternative, utilizing sparse, event-driven communication to process information~\cite{roy2019towards}. SNNs have demonstrated energy efficiency advantages due to their inherent sparsity and temporal dynamics. However, their efficiency gains has come at the cost of performance degradation. SNNs are limited in their adoption as they do not match the performance of traditional Artificial or Recurrent Neural Networks, and the performance gap widens as task complexity increases. This stems from difficulties in training nonlinear sequence-based models, extreme low-precision activations, and specialized hardware requirements~\cite{pfeiffer2018deep}.

This work presents a unified solution to both problems: (i) obtaining the data communication benefits of SNNs in a mode where, (ii) they can algorithmically scale up to more complex domains. We propose an algorithm-architectural co-design approach to heterogeneous neural networks (HNNs). In doing so, we reduce data transfer overhead and energy consumption across bandlimited regions of the system while preserving ANN computations within each chip to maintain high accuracy, ultimately drawing upon the strengths of both SNNs and ANNs.

Earlier research has explored hybrid models that interleave spiking layers with traditional non-spiking layers to enhance deep learning performance~\cite{liu2024advancing, fang2021deep, voelker2020spike, seekings2024towards}. However, they have not yet demonstrated how strategically partitioning spiking layers around bandwidth bottlenecks can amplify these gains. This study is the first to systematically align spike-based processing with resource-constrained die-to-die interfaces and show consistent performance improvements across all tested domains, from language processing to computer vision. Comparisons between ANNs, SNNs, and HNNs, show that HNNs operate at a Pareto-optimal point for accuracy, throughput, and efficiency metrics.

\textbf{\textit{The key contributions of this paper are:}}

\begin{enumerate}
    \item A systematic method to partitioning neural networks into SNN and ANN components when adapting modern deep learning architectures into HNNs based on bandwidth-constrained communication regions, reaching accuracy at least on-par with sole ANN and recurrent neural network (RNN) counterparts.
    \item A hybrid inter-die architecture that integrates spiking and non-spiking processing elements, with a flexible interconnect fabric for routing mixed spike and dense data packets, optimized for HNN workloads, achieving up to 15.2$\times$ speedup compared to conventional ANN implementations, and 5.3$\times$ energy efficiency improvements.
    \item A Network-on-Chip (NoC) simulation framework capable of analyzing latency, throughput, and energy consumption for ANN, SNN, and HNN models. It calculates energy estimates for Extended-Mux I/O (EMIO), Memory (MEM), Processing Element (PE), and Router components and provides extensibility for custom layer definitions, enabling detailed performance modeling and architectural exploration.
\end{enumerate}

\begin{figure*}[!h]
    \centering
    \includegraphics[width=0.9\textwidth]{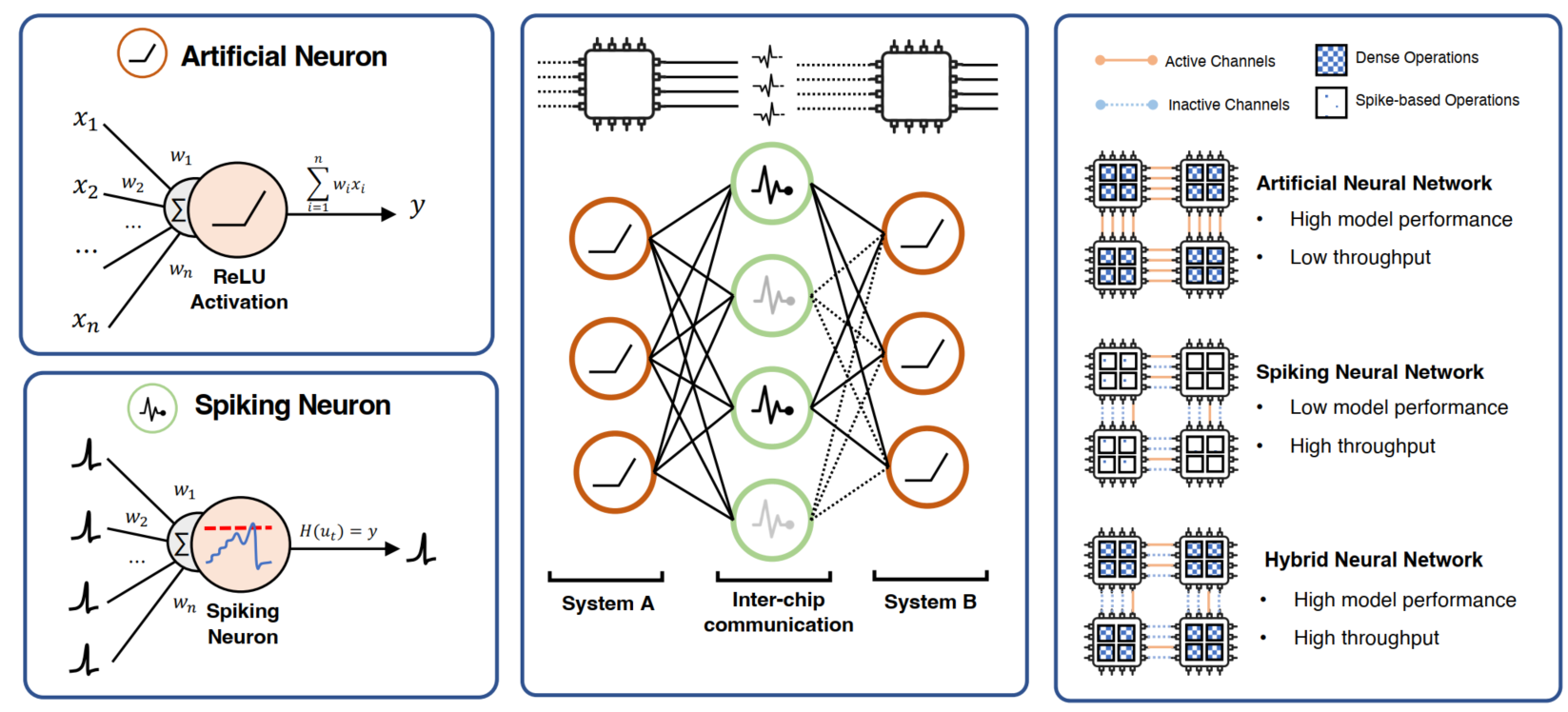}
    \caption{An overview. Artificial neurons are integrated with spiking neurons, where $x_i$ denotes an input, $w_i$ denotes a weight, $H(\cdot)$ is a thresholding function, $u_t$ is the membrane potential of the spiking neuron at time $t$. Spiking neurons are positions at the peripheral of demarcation boundaries between systems (inter-chip communication is used in this paper). This enables sparse data communication using spike-based die-to-die interfaces, with dense ANN operations within cores.}
    \label{fig:intro}
\end{figure*}

The remainder of this paper is organized as follows: Section~\ref{sec:related} provides background on the challenges of inter-chip communication and reviews related work. Section~\ref{sec:arch} details the algorithm-architecture co-design methodology and its various components. Section~\ref{sec:method} describes our experimental setup, benchmark models, and the custom Network-on-Chip simulation framework developed for evaluation. Section~\ref{sec:results} presents and discusses our results. Finally, \ref{sec:conclusion} concludes the paper with related works and an outline of future directions.


\section{Related Work and Motivation}
\label{sec:related}

\subsection{The Challenge of Scaling SNNs}
\textbf{Background:} ANNs and SNNs can model similar network topologies but differ in their neuron models. Artificial neurons integrate simultaneous inputs, whereas SNNs use spiking neurons, which act as binarized, linear recurrent units updating only when their membrane potential $U$ crosses threshold $\theta$. This behavior is captured by the leaky integrate-and-fire (LIF) model: 

\begin{align}
    \tau_m \frac{dU}{dt} = -U + RI \implies U_{t+1} = \beta U_t + (1-\beta)I_t,
\end{align}

\noindent where $\beta = e^{-\Delta t / \tau_m}$ and $I_t = \sum^n_{i=1} w_ix_{i,t}$ is the weighted sum of inputs. The left-hand side expresses continuous-time dynamics, while the discrete-time equation on the right is suited to hardware implementations~\cite{pedersen2024neuromorphic}. A spike is emitted if $U_t \ge \theta$. For a detailed derivation, see Ref.~\cite{eshraghian2023training}.

\textbf{The Current State of Large-Scale SNNs:} The past decade of SNN research has shifted from modeling strictly biological features to applying deep learning methods at scale. Yet, large-scale success remains limited compared to conventional deep learning. For instance, `Spikformer'~\cite{zhou2024spikformer} and `Spike-driven Transformer'~\cite{yao2023spike} adapt LIF neurons for self-attention on ImageNet, but have yet to achieve language generation. Spikformer-v2 (51M parameters) reaches a top-1 accuracy of 80.38\% on ImageNet, while a similarly sized EfficientNet (54M) reaches 86.2\%~\cite{tan2021efficientnetv2}. 

\noindent SpikeGPT, which relies on a more complex RWKV-based neuronal formulation, remains the largest SNN to demonstrate language modeling, although performance gains of its 1-B parameter variant were only marginally better than those of a 216-M parameter version~\cite{zhu2023spikegpt, peng2023rwkv}. Despite their potential efficiency improvements, current evidence indicates that SNNs often saturate faster than non-spiking networks. In order of importance, the reasons are:
\begin{itemize}
    \item significant information loss in binarized activations,
    \item slow training due to temporal nonlinearities that impede parallelization,
    \item non-differentiable operators, although this is largely resolved by surrogate gradient methods.
\end{itemize}

We hypothesize that purely spiking architectures offer diminishing returns where bandwidth is abundant and deterministic runtimes are desirable. Although SNNs reduce communication overhead with event-driven processing, their benefits diminish when most data transfers occur on chip. Hence, we propose to confine spiking layers to bandwidth-constrained regions while preserving dense, procedural processing elsewhere.

\subsection{Heterogeneous Neural Networks}
\textbf{Enhancing Performance at the Cost of Complexity:} Incorporating richer neuronal dynamics has improved domain-specific performance, as evidenced by language generation with many linear state-space models~\cite{gu2023mamba, zhu2024scalable} and their variants in SNNs~\cite{voelker2019legendre, stan2024learning, bal2024rethinking}. However, running dense layers on hardware designed for sparse operations leads to significant overhead in throughput and energy efficiency. Even sophisticated spiking neuron models, when universally applied, often underperform on tasks that demand ample on-chip bandwidth or deterministic runtimes. This observation motivates a design that only leverages the sparseness benefits of SNNs in bandwidth-limited regions, reserving dense computations for more conventional hardware.

\textbf{Heterogeneous Architectures:} Merging SNNs and ANNs has been attempted in various forms, typically to mitigate the inherent performance drops seen in purely spiking networks. For instance, non-spiking interneurons have augmented SNNs for improved fidelity in robotics~\cite{liu2024advancing}, and hybrid modes balancing between single-bit spike-based and full-precision behavior have been explored~\cite{voelker2020spike,seekings2024towards}. Sometimes, an ANN read-out is used to circumvent non-differentiability in SNN layers~\cite{fang2021deep}. Ref.~\cite{chang202373} represents one of the few hardware demos running both SNN and ANN concurrently for real-time tasks, though the two networks remain distinct. 

Despite these efforts, existing hybrid approaches often overlook the added complexity that dense layers impose on spiking hardware while also risking performance degradation from the spiking layers when viewed from a conventional deep learning standpoint. The result is a compromise in which neither SNN nor ANN performance genuinely benefits from their integration. We argue that a carefully partitioned heterogeneous approach, structured around bandwidth and energy constraints, can maintain high performance while scaling more efficiently than current all-spiking or all-dense solutions.




\subsection{Accelerators for Heterogeneous Workloads}

Modern SNN accelerators, such as Loihi 2~\cite{loihi2}, TrueNorth~\cite{TrueNorth}, ReckOn~\cite{ReckOn}, and others~\cite{ODIN,narayanan2020spinalflow,painkras2013spinnaker}, demonstrate substantial energy efficiency and can exceed 1000$\times$ improvements over conventional ANN systems~\cite{loihi2}. While SNNs performance generally lag behind deep learning, ANN-centric accelerators~\cite{eyerissv2,NeuronLink,gao2017tetris} excel at large-scale throughput but struggle with sparse computation. Hybrid solutions, such as NEBULA~\cite{singh2020nebula} and others~\cite{dampfhoffer2022snns}, combine ANN and SNN cores in a single chip, yet typically ignore how dense layers compromise throughput on spiking hardware or how spiking layers reduce ANN performance.

Our proposed co-design approach to HNNs places spiking layers at bandwidth-constrained chip boundaries and retains dense layers on more conventional processing elements. This leverages event-driven communication for high-latency inter-chip links while exploiting parallel, deterministic ANN workloads within each die. Integrating sparse and dense processing in a single platform addresses scalability hurdles, particularly bandwidth and energy constraints, ultimately preserving high accuracy while substantially reducing overall power and latency across diverse deep learning tasks.

While many ANN accelerators focus on extreme reconfigurability of the Network-on-Chip (NoC) fabric~\cite{lu2021runtime}, their optimization is often task-specific. We address these challenges by offering a unified platform capable of custom DNN mapping across both SNN and ANN layers with co-designed interfaces across layers, achieving greater flexibility and efficiency in hybrid workloads. This approach provides a versatile way to scale large DNNs with reduced energy and latency overhead, selectively harnessing SNNs where their advantages are most pronounced.

\section{SW/HW Co-Design of HNNs}
\label{sec:arch}

In this section, we present our inter-die architecture that combines spiking neurons for communication and artificial neurons for computation, effectively addressing bandwidth bottlenecks in die-to-die communication. The data exchanged between chips is compressed into a sparse representation, where the sparse transformation is learned using gradient descent during training time, rather than routing pre-defined signals, cutting down on overhead while preserving computational fidelity.

We evaluate this heterogeneous approach across multiple datasets, models, and across various levels of activation sparsity, comparing its performance to representative ANN and SNN baselines. Our approach demonstrates superior efficiency and scalability while remaining adaptable to alternative core designs. Detailed results, including metrics on energy consumption and latency, are presented in Section~\ref{sec:results}.

\subsection{Proposed Architecture}
\begin{figure*}[!h]
    \centering
    \includegraphics[width=\linewidth]{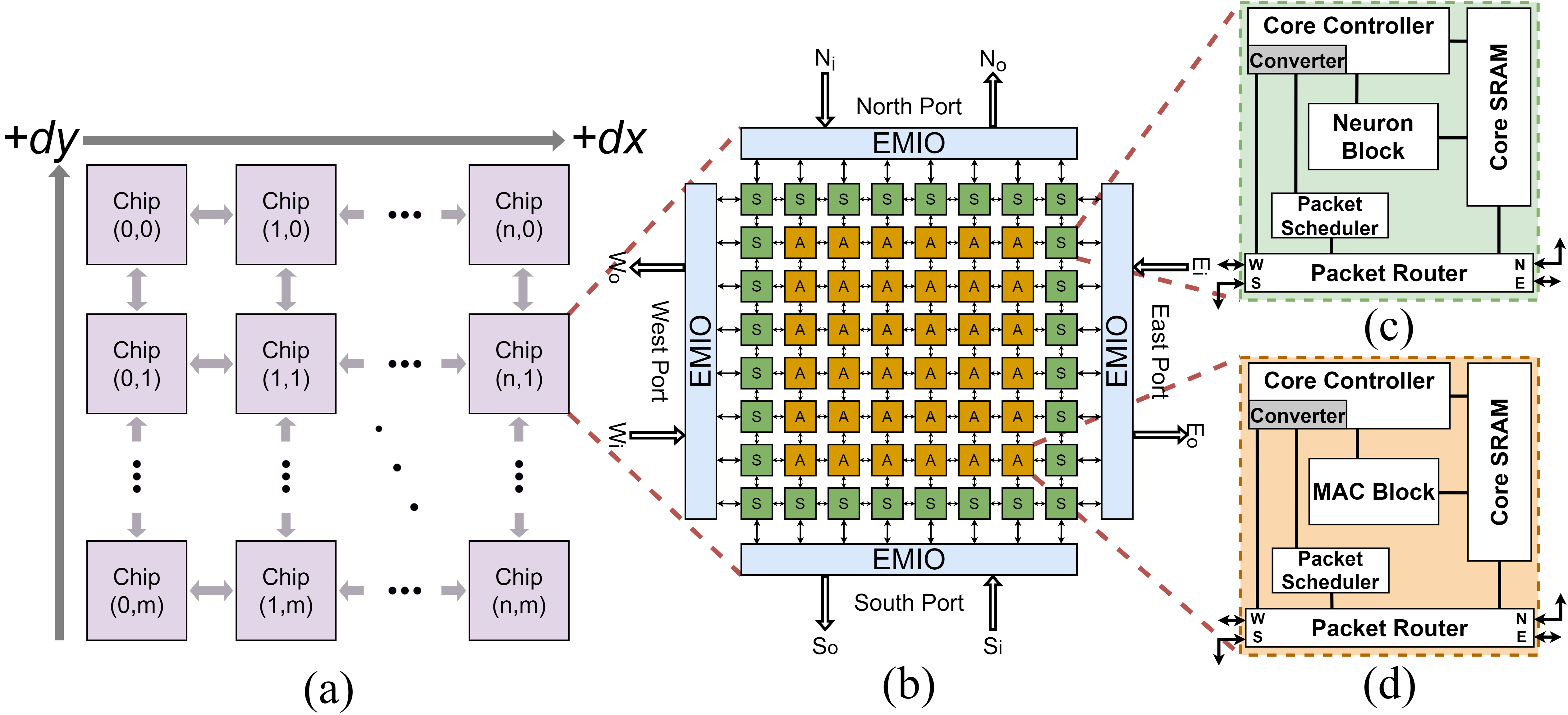}
    \caption{2-D Mesh NoC Hybrid Hardware Accelerator Overview. A high-level overview of the hybrid approach to solve the limitations of pure ANN or SNN accelerators. \textbf{(a)} Proposed Hybrid Architecture 2-D inter-die processing array. \textbf{(b)} Architecture overview showing the 2-D Mesh NoC SNN peripheral cores and ANN interior core grid with two unidirectional ports on each side connected through the EMIO. \textbf{(c)} SNN Core with proposed hybrid architecture's CLP converter. \textbf{(d)} ANN Core with proposed hybrid architecture's converter.}
    \label{fig:arch_overview}
\end{figure*}

To leverage the complementary benefits of SNN and ANN designs in a unified acceleration system, our architecture integrates a 2-D Mesh Network-on-Chip (NoC) with an $8\times 8$ grid of Core Tiles. Key components include an Extended Multiplexer IO (EMIO) block that interfaces with other chips/die and a Computational Cross-layer Packet (CLP) converter connected to each core controller, as depicted in Fig.~\ref{fig:arch_overview}. This CLP converter enables data conversion between spiking and artificial layers, ensuring efficient inter-layer communication across the system.


\subsection{2-D Mesh-based NoC Grid Design} 
We employ a NoC-based 2-D mesh design with static X-Y routing due to its scalability, reconfigurability, and simplicity. The NoC fabric consists of 28 spiking cores located at chip boundaries and 36 artificial cores centered within the chip. Both core types are designed to share common architecture blocks, with optimizations tailored to their respective neuron computations. Each core comprises five primary design blocks, as illustrated in Fig.~\ref{fig:arch_overview}.

Packet communication across cores is facilitated by deterministic X-Y routing handled by a Packet Router. The \textit{dx} and \textit{dy} fields within the packet header (Tab.~\ref{tab:packet_struct_params}) enable packets to traverse up to 256 cores before reaching a network-mapping repeater core for further routing. This configuration supports inter-chip communication across up to eight chips in any direction, with repeater cores extending communication as needed. To prevent deadlock, routing prioritizes the X (East/West) direction~\cite{TrueNorth}.

Table~\ref{tab:arch_params} provides detailed on-chip SRAM specifications for both ANN and SNN cores. The supply voltage for the architecture is set to the minimum supported by the 65nm process node library, ensuring efficient operation while maintaining compatibility with the technology node.

\begin{table}[!h]
    \centering
    \caption{Architectural Parameters}
    \label{tab:arch_params}
    \begin{tabular}{c c c c}
        \hline
        \textbf{Parameter} & \textbf{ANN} & \textbf{SNN} & \textbf{HNN}  \\
        \hline
        \# Spiking Cores
        & - & 64 Cores & 28 Cores \\
        \# Artificial Cores
        & 64 Cores & - & 36 Cores \\
        NoC frequency 
        & 200 MHz & 200 MHz & 200 MHz \\
        Supply voltage
        & 1.0V & 1.0V & 1.0V \\
        On-Chip SRAM & 1.1 MB & 860 KB& 1 MB\\
        \hline
    \end{tabular}
\end{table}

\subsection{ANN \& SNN Core Designs}

\begin{table}[!h]
    \centering
    \caption{Comparison of ANN and SNN Core Parameters}
    \label{tab:combined_core_params}
    \begin{tabular}{c c c}
        \hline
        \textbf{Parameters} & \textbf{ANN} & \textbf{SNN} \\
        \hline
        \# neurons / \# axons & 256 / 256 & 256 / 256 \\
        \# synapses & 64k & 64k \\
        \hline
        core SRAM & 13.75 KB & 12.93 KB \\
        scheduler SRAM & 4 KB & 0.5 KB \\
        \hline
        MAC precision & 8b $\times$ 8b & -- \\
        accumulator precision & 32b & -- \\
        spike precision & -- & 1b \\
        \hline
        weight / neuron potential precision & 32b & 8b \\
        activation precision & 8b & -- \\
        \hline
    \end{tabular}
\end{table}

Our HNN approach incorporates both ANN and SNN cores, each designed as a synchronous, clock-driven module. While the core architectures are distinct in their components, they adhere to the core design principles established by platforms such as RANC and TrueNorth~\cite{RANC, TrueNorth}, with enhancements inspired by Eyerissv2~\cite{eyerissv2}. To maintain a homogeneous core array within the HNN accelerator, changes between ANN and SNN cores were minimized, focusing only on the differences essential for their respective computations. This streamlined, reconfigurable design allows for meaningful comparisons across ANN, SNN, and HNN accelerators, emphasizing performance and efficiency improvements in our benchmarks. Core parameters for ANN and SNN implementations are detailed in Table~\ref{tab:combined_core_params}, respectively.

Each core comprises a processing element (PE), a packet scheduler, and SRAM memory for storing weights and activations. A single core accommodates 256 neurons and axons, resulting in a system-wide total of over 64k synapses. This synapse count represents the maximum capacity of the core grid, which places constraints on large-scale DNN implementations. Models with significantly higher synapse counts per layer, such as those with dense fully connected layers, must map connections across multiple hardware iterations. For instance, two fully connected layers of 256 neurons fully utilize the available synapse capacity. This highlights the importance of efficient resource allocation and connection mapping to enable the deployment of deeper and more complex networks within the hardware's architectural limits.

The multiply-and-accumulate (MAC) units and accumulators within the PE of the ANN core are designed with 32-bit weights, 8-bit activations, an 8-bit$\times$8-bit MAC, and 32-bit accumulators. In contrast, the SNN core PE features 8-bit weights, 8-bit membrane potentials, and 1-bit spikes. These differences in bit width and precision influence the design of the scheduler and core SRAM, with values scaled accordingly to fit their respective requirements. To minimize weight movement and maximize SRAM reuse, both ANN and SNN cores employ a weight-stationary dataflow, where weights and membrane potentials remain fixed in local core memory while activations and spikes flow through the PEs.

For the SNN core scheduler, the SRAM consists of 16$\times$256-bit entries to manage the 256 axons with a maximum delay of 16 ticks. This design enables temporal spike organization across all axons with a memory footprint of 0.5~KB per SNN core scheduler. In comparison, the ANN core scheduler SRAM is extended to accommodate the higher 8-bit activation precision, with 16$\times$2048-bit entries totaling 4~KB per ANN core scheduler.

Similarly, the core SRAM is tailored to the requirements of each network. The SNN core features 256 $\times$ 410-bit entries, providing 12.93~KB of SRAM per core. These entries include synaptic connections, weights, potentials, neuron parameters (256 bits), packet destinations (124 bits), and delivery ticks (4 bits). For the ANN core, the SRAM is expanded to 256 $\times$ 440-bit entries, totaling 13.75~KB per core. This additional capacity accommodates the larger activation precision and associated parameters.

The Core Controller orchestrates the control path signals and is programmable to handle various layer types and computations. It interfaces with all core components, ensuring efficient data processing and coordination across the system.
Specifically for SNN layer data communication, static dataset inputs must be encoded with multiple timesteps (as a buffered train of spikes) to encode information within the temporal domain. The 4-bit delivery time mentioned above has the ability to encode up to a 16-step spiking input. Utilizing higher time-steps in inputs increase the number of operations in the PE for a given spiking layer. Within dynamic datasets, we can utilize the temporal aspects of the inputs and do not need to encode the inputs across multiple timesteps.

\subsection{EMIO Block Design}

The Extended Mux I/O (EMIO) is a low-overhead interconnect designed to optimize die-to-die packet communication. This approach adapts elements from the TrueNorth interconnect design~\cite{TrueNorth}, scaling its merge-split block, originally interfacing 32 ports—to fit the 8 boundary spiking cores in our architecture. At the chip boundary, the NoC includes 64 unidirectional ports (32 input, 32 output) multiplexed down to 8 unidirectional ports at the I/O pads, enabling efficient communication while minimizing resource usage.

\begin{figure}[!h]
    \centering \includegraphics[width=\columnwidth]{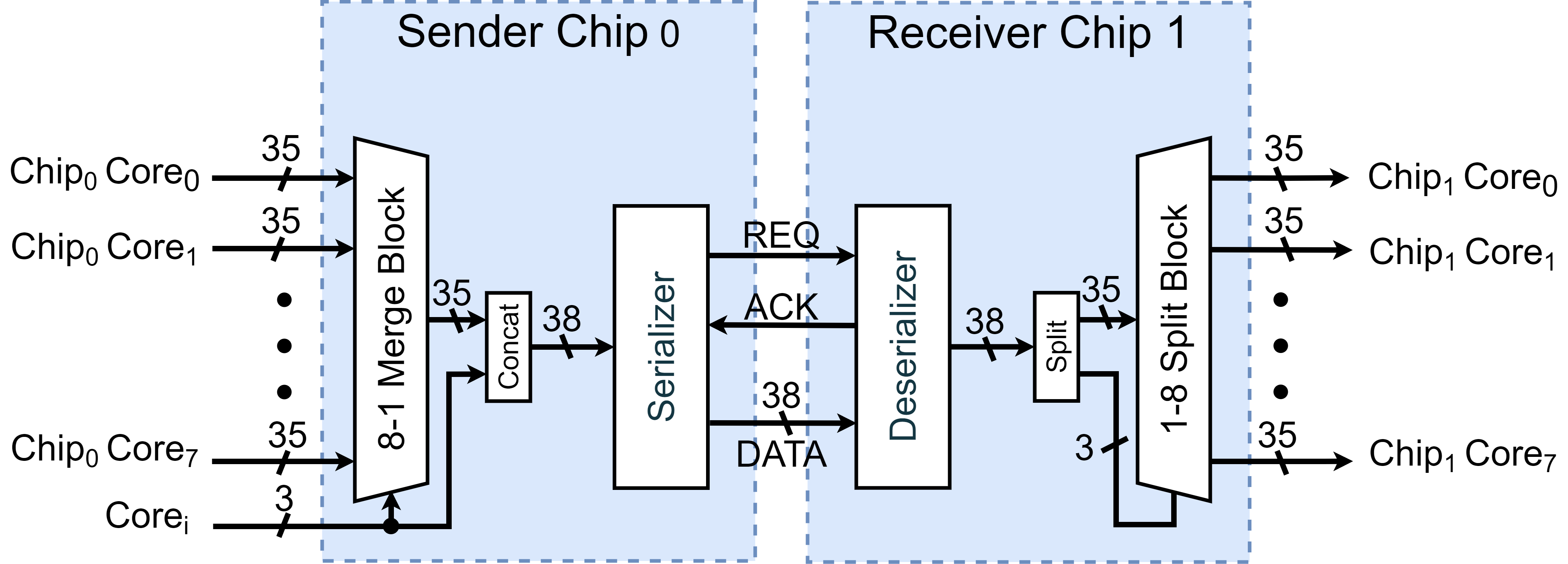}
    \caption{High Level Interconnect (single port). Flow of I/O at a high level. Asynchronous FIFO Buffers within the Merge and Split Blocks control incoming Spike Packets from the 8 peripheral cores \& outgoing to the mapped 8 next chip peripheral cores through SerDes block. 
    }
    \label{fig:EMIO}
\end{figure}

TrueNorth's design exhibited a significant reduction in spike bandwidth—over 640$\times$ lower at the chip boundary compared to on-chip communication. This was caused by 2$\times$ serialization, 32-to-1 port multiplexing, and a 10$\times$ disparity between the NoC and I/O pad clock frequencies. Unlike TrueNorth's asynchronous design, our EMIO employs a synchronous 200 MHz clock frequency at the chip boundary, eliminating clock disparities. As shown in Fig.~\ref{fig:EMIO}, our design uses 8-to-1 port multiplexing with reduced serialization. A synthesized RTL implementation of our EMIO design revealed a die-to-die latency of 76 clock cycles for a single packet, with the serialization stage accounting for 38 cycles. Importantly, the serialization process occurs in parallel across the 8 peripheral ports connected to the boundary cores.

To support die-to-die communication, packets from the 8 peripheral cores are processed through a SerDes for transmission via the I/O pads. Each 35-bit packet is tagged with 3 additional bits for origin/destination mapping, resulting in a compact 38-bit format. The SerDes ensures efficient serialization, minimizing communication overhead and enabling precise packet routing across die-to-die interfaces. This design balances compactness and performance for high-throughput communication.

\begin{figure}[!h]
    \centering
    \begin{subfigure}[!h]{\linewidth}
        \centering
        \includegraphics[width=\linewidth]{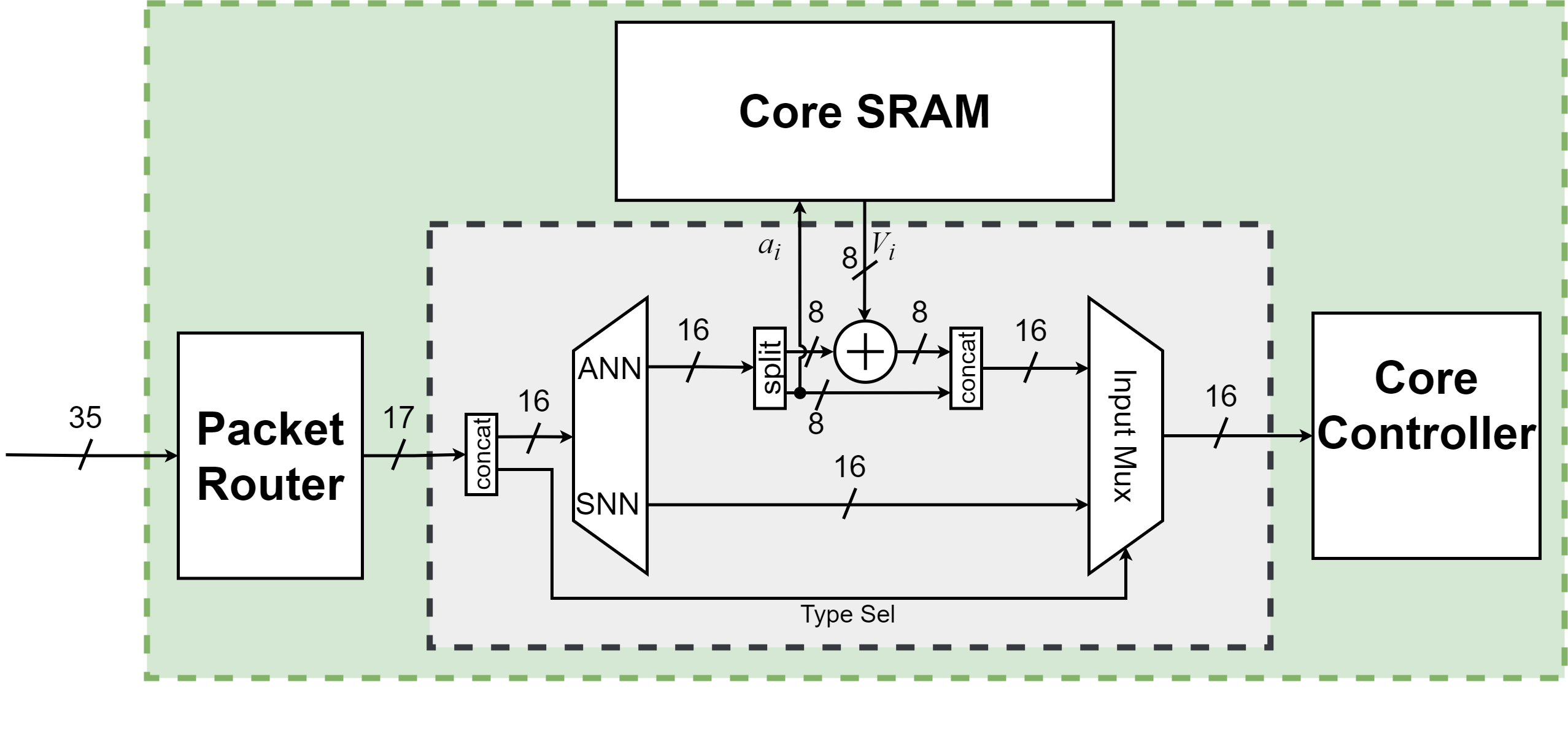}
        \caption{Activations are accumulated with the axon connection index within the core SRAM and are treated like incoming spikes.}
        \label{subfig:cross_converter_a}
        \vspace{5mm}
    \end{subfigure}
    \hfill
    \begin{subfigure}[!h]{\linewidth}
        \centering
        \includegraphics[width=\linewidth]{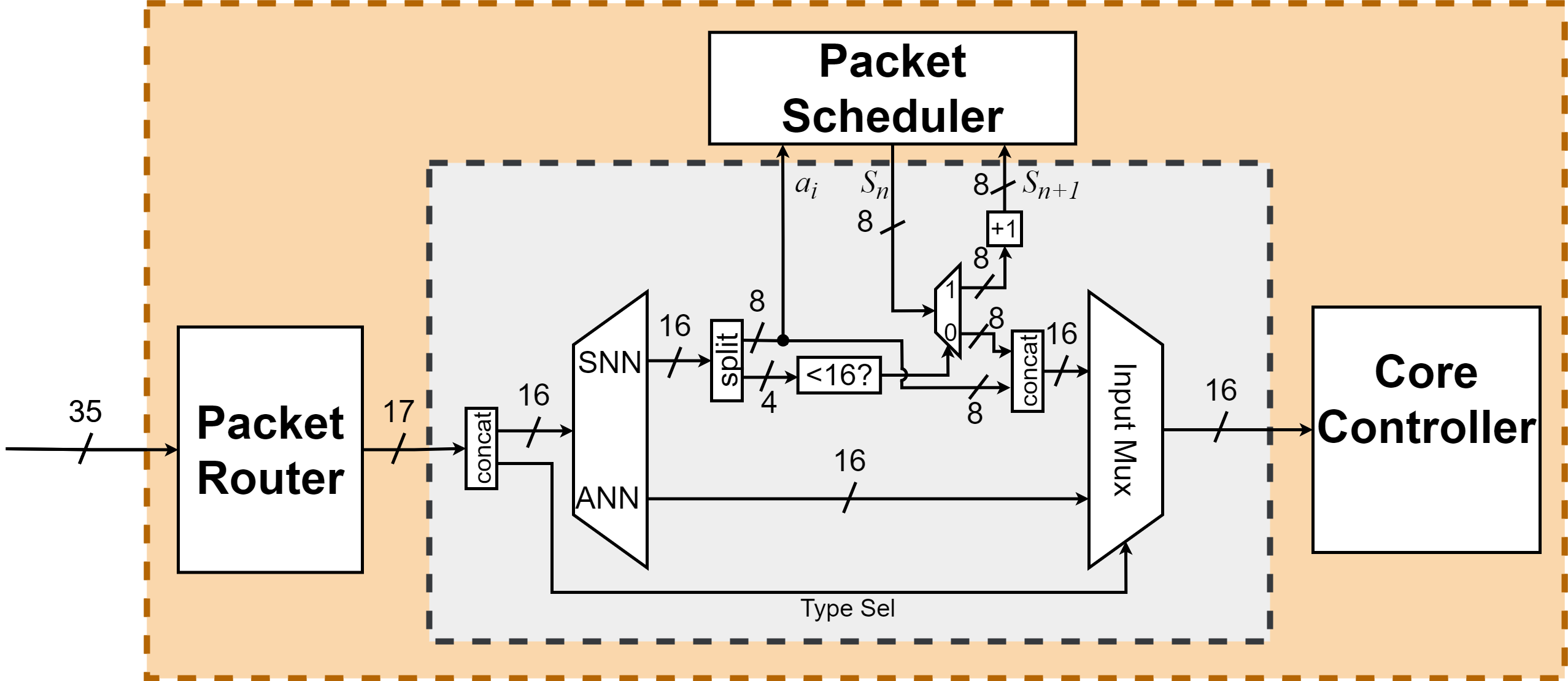}
        \caption{Spikes accumulated for set max tick delay (16). The number of spikes is stored within the scheduler as an 8-bit value, transforming into activation.}
        \label{subfig:cross_converter_b}
        \vspace{5mm}
    \end{subfigure}
    \caption{Cross-Layer Activation-to-Spiking and Spiking-to-Activation Packet Converter Design.}
    \label{fig:cross_converter}
\end{figure}

\subsection{CLP Converter Design}
For packet conversion, the core design incorporates a CLP converter for each of their 256 axons. This converter translates rate-encoded spike trains into activation-encoded artificial packets and vice versa. The logical design, depicted in Fig.~\ref{fig:cross_converter}, selects the packet type based on the packet type field in Table~\ref{tab:packet_struct_params} to identify the packet type and perform in-line packet conversion with minimal overhead.

\begin{table}[!h]
    \centering
    \caption{Packet Structure Parameters}
    \label{tab:packet_struct_params}
    \begin{tabular}{c c c}
        \hline
        \textbf{Field} & \textbf{ANN} & \textbf{SNN} \\
        \hline
        dx core dest. & {9 bits} & {9 bits} \\
        dy core dest. & {9 bits} & {9 bits} \\
        type & {1 bit} & {1 bit} \\
        axon index & {8 bits} & {8 bits} \\
        \hline
        Payload & 8-bit & 4-bit + padding \\
        \hline
    \end{tabular}
\end{table}

For activation-to-spiking packet conversion (Fig.~\ref{subfig:cross_converter_a}), the CLP converter accesses the spiking neuron's potential and directly accumulates the activation value. This process generates a rate-encoded spike sequence proportional to the activation value, \( a_i \in [0, 2^b - 1] \), distributed across a tick window of size \( T \). The spike train \( s_i(t) \in \{0, 1\} \) is generated according to the deterministic rate coding rule:

\begin{equation}
    \label{eq:act_to_spike}
    s_i(t) = 
    \begin{cases}
    
        1 & \text{if } t < \left\lfloor \dfrac{a_i}{T} \right\rfloor \\
        0 & \text{otherwise}
    \end{cases}, \quad t \in [0, T-1]
\end{equation}

For spiking-to-activation packet conversion (Fig.~\ref{subfig:cross_converter_b}), the CLP converter accumulates the incoming spikes over a predefined maximum tick delay \( T \), storing intermediate values in the Packet Scheduler. After accumulation, the total spike count \( S_i = \sum_{t=0}^{T-1} s_i(t) \) for axon \( i \) is scaled into an activation value using the inverse mapping:

\begin{equation}
    \label{eq:spike_to_act}
    a_i = \left\lfloor \dfrac{2^b - 1}{T} \cdot \sum_{t=0}^{T-1} s_i(t) \right\rfloor
\end{equation}

This bidirectional conversion enables efficient and unified processing of both ANN-style and SNN-style data representations within the same core tiled NoC architecture, supporting flexible hybrid neural network computations.

\section{Methodology}
\label{sec:method}

\subsection{Benchmark Models}

\begin{figure}[!h]
    \centering
    \includegraphics[width=\linewidth]{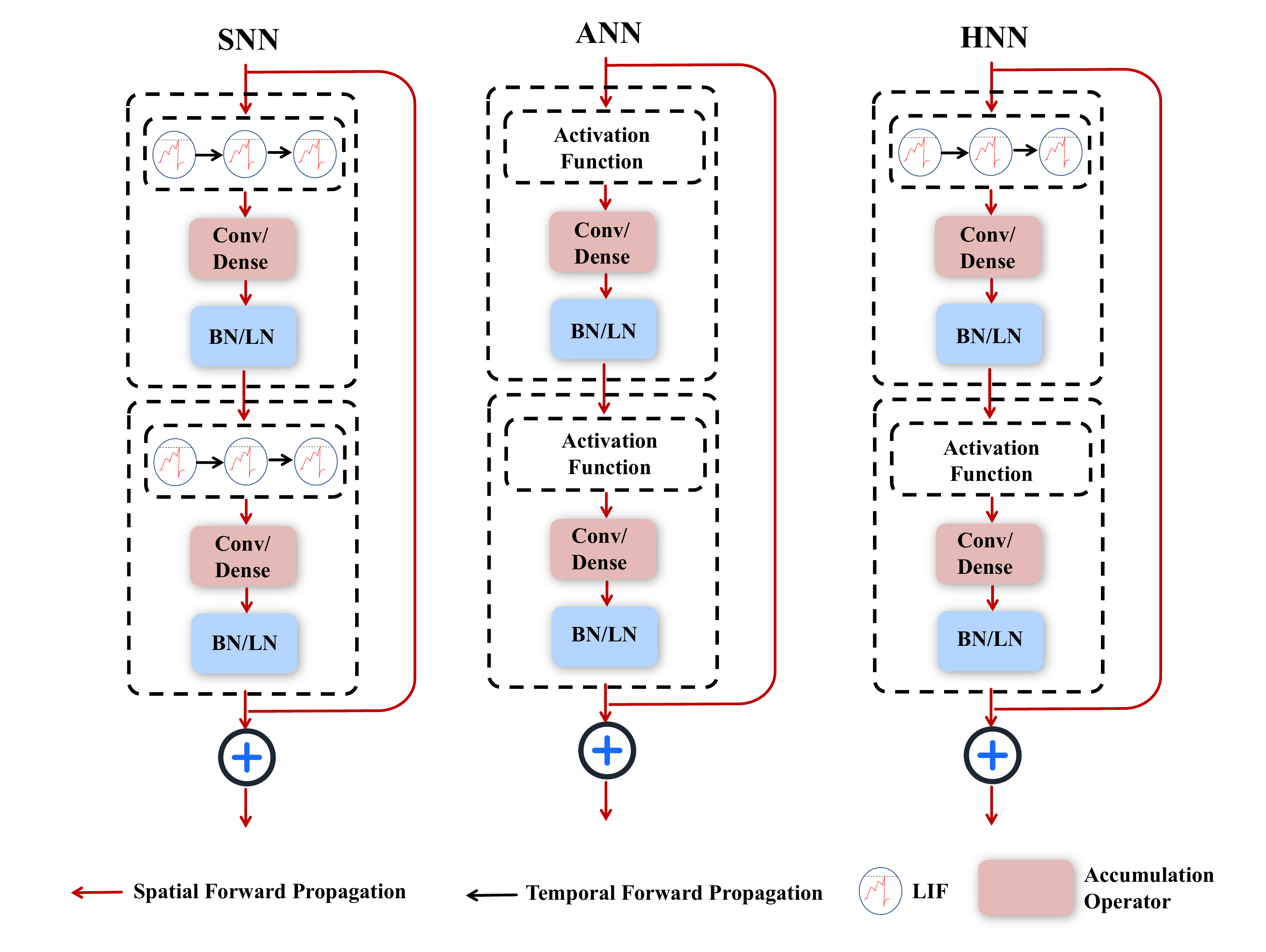}
    \caption{Comparative Analysis of MS-ResNet Architectures: This diagram illustrates three distinct MS-ResNet architectures, highlighting the utilization of Batch Normalization (BN) and Layer Normalization (LN). BN is predominantly employed in computer vision tasks, while LN is more suited for language modeling tasks. The architecture variation extends to the layer types; convolutional layers (Conv) are used in computer vision tasks, whereas dense layers are more common for language modeling.}
    \label{fig:resnet}
\end{figure}

We integrate MS-ResNet~\cite{msresnet}, EfficientNet-B4~\cite{tan2019efficientnet}, and RWKV~\cite{peng2023rwkv} as representative architectures for computer vision and language models, respectively, to evaluate performance across diverse and challenging datasets. MS-ResNet18, a spike-driven variant of ResNet, is widely used in contemporary SNN research~\cite{msresnet1,msresnet2,msresnet3} and is inherently compatible with both SNN and ANN architectures. The specific configuration of MS-ResNet18 is illustrated in Fig.~\ref{fig:resnet}. Unlike standard ResNet models, MS-ResNet18 employs membrane potential summation, enabling efficient spike-driven operations by modulating membrane potentials. For HNN-compatibility, each block uses LIF neurons, while inter-block connections maintain ANN compatibility, ensuring optimized input-output performance.

For language modeling, we leverage RWKV, a recurrent neural network (RNN)-based model designed as a lightweight alternative to transformers for language generation. RNN-based models like RWKV are particularly well-suited for integration with SNNs due to their inherent temporal processing capabilities. RWKV achieves competitive results compared to transformer-based architectures, with significantly lower computational overhead. In our implementation, we incorporate a residual connection similar to that used in MS-ResNet.

\subsection{Simulation Environments}

We developed a custom simulation environment for multi-chip accelerator systems to evaluate artificial, spiking, and hybrid networks with a focus on die-to-die communication. While existing simulators such as RANC and NN-Noxim~\cite{RANC,NN-Noxim} provide valuable tools for network evaluation, they are not specifically designed for modeling multi-chip, partitioned hybrid networks. To address this gap, we created a high-level simulation framework tailored for large-scale hybrid DNN models, offering flexibility for adding custom DNN layer definitions for SNNs, ANNs, and HNNs.

\begin{figure}[!h]
    \centering
    \includegraphics[width=\columnwidth]{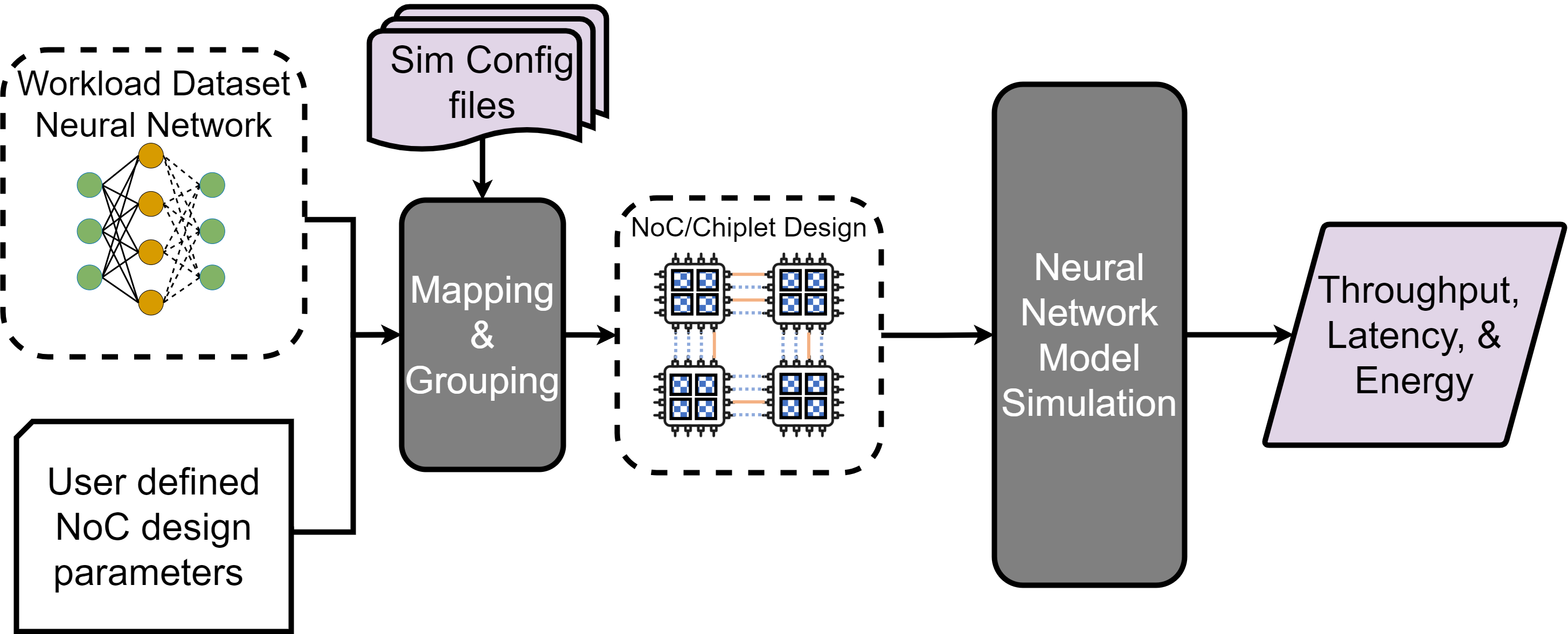}
    \caption{Model simulation workflow. Visualizing how ANN, SNN, \& HNN Neural Networks are mapped onto the NoC Designs and how all the performance metrics, throughput, latency, and energy are calculated. 
    }
    \label{fig:sim_workflow}
\end{figure}

The simulation workflow, illustrated in Fig.~\ref{fig:sim_workflow}, employs a co-designed approach where user-defined network workloads and NoC parameters are mapped onto the hardware. This enables layer-accurate simulation of large hybrid DNNs on custom architectures. The simulator adheres to a synchronous, clock-driven design for all ANN, SNN, and HNN components, mirroring the architectural behavior of the proposed system.

Energy and power calculations are based on the ORION 2.0 methodology~\cite{orion2}, which models Intel's 65nm 80-core chip~\cite{intel80core}. The parameters were scaled for our design's 1.0~V core supply voltage and 200~MHz NoC frequency, ensuring realistic energy efficiency results. Key simulation metrics include operations per layer, routed packets, and local packets. Operations correspond to MAC operations in ANN layers and accumulate (ACC) operations in SNN layers. Routed packets represent inter-core communication, calculated based on the average number of hops per packet using directional-X neural network mapping and X-Y packet routing equations (\ref{eq:average_hops}) and (\ref{eq:routed_packet_calc}). Local packets represent intra-core communication sent through the local port to the processing element (PE) for computation, providing a comprehensive evaluation of computational and communication overheads.

\begin{equation}
    \label{eq:average_hops}
    Average Hops = \left| \textit{M}_{L_{i-1}} - \textit{M}_{L_{i}} \right| + 1
\end{equation}

Where \textit{M}$_{L_{i-1}}$ is the middle core coordinates of the current layer and \textit{M}$_{L_{i}}$ is the middle core coordinates of the previous layer. The absolute Manhattan distance between layer midpoints approximates the average hops taken by a routed packet.

\begin{equation}
    \label{eq:routed_packet_calc}
    Routed Packets = Average Hops \times  Local Packets
\end{equation}

Calculating total number of routed packets in~(\ref{eq:routed_packet_calc}) is the number of local packets received through the local port multiplied by the number of average hops calculated in~(\ref{eq:average_hops}).

The scalable NoC simulators enabled us to model large, deep networks, including fully connected layers with over 19 million parameters. The three models—RWKV, MS-ResNet18, and EfficientNet-B4—were simulated with approximate workloads to evaluate system performance. For SNN models, we assumed a high input sparsity of 90\% (10\% spiking activity) and rate-encoded dataset inputs with a temporal window of $T$ = 8 time steps. This level of sparsity aligns with methodologies used in SNN accelerators such as SpinalFlow~\cite{narayanan2020spinalflow,lemaire2022analytical}.

Following established methodologies discussed in prior work~\cite{DNNoC-Sim,lemaire2022analytical}, we calculated the computational workload for both SNN and ANN layers, including convolutional, depthwise-convolutional, and pooling operations. For ANN models, operations were measured in terms of MAC, while for SNNs, operations were quantified as ACC counts, capturing the distinct computational requirements of each network type. HNN models are partitioned using ANN-style activation-based data representations for intra-chip communication and interior core computations (MACs), and SNN-style spike-based representations for inter-chip communication and peripheral core computations (ACCs). This partitioning methodology leads to simulation results that reflect mixed traffic patterns and computational workloads representative of hybrid neural networks.

\begin{figure*}
     \centering
    \includegraphics[width=1\linewidth]{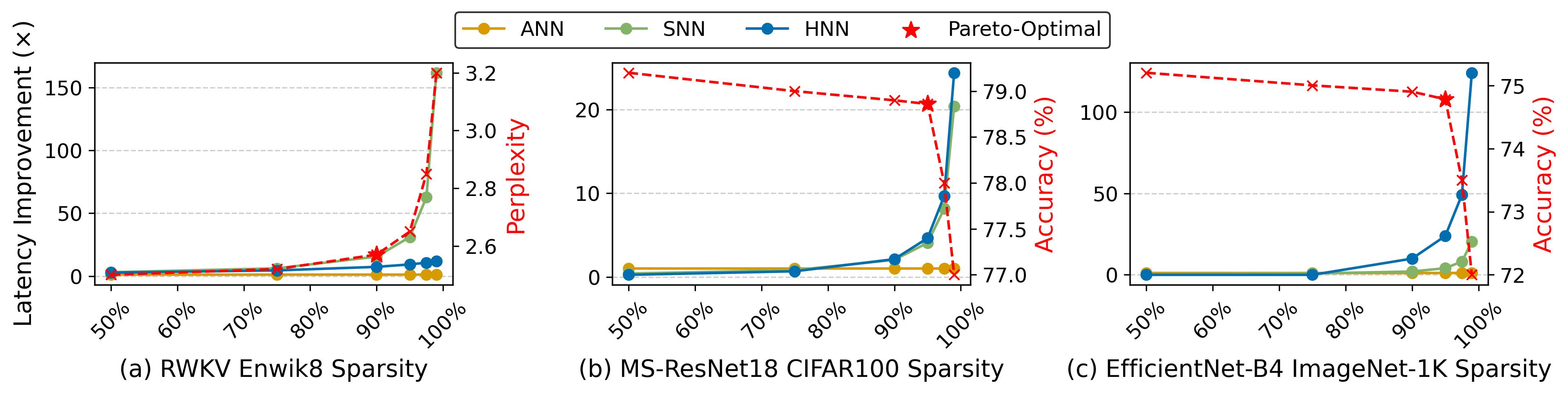}
    \caption{Activation sparsity sweep for each model. Latency improves with more sparsity. Model performance (perplexity/accuracy) is generally stable until a phase transition with excessive sparsity (beyond 95\% for RWKV, and beyond 97.5\% for the computer vision tasks.}
    \label{fig:sparsity-sweep}
\end{figure*}

\begin{figure*}
     \centering
    \includegraphics[width=0.9\linewidth]{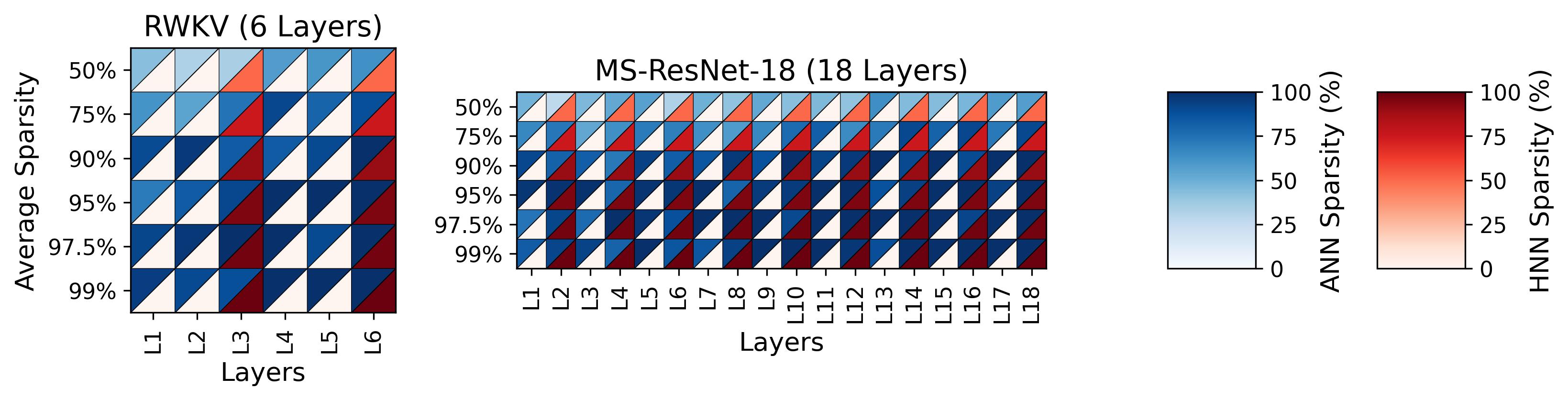}
    \caption{Activation heatmap across layers. HNNs are only sparsified at the spiking boundary layers, and partitioned based on the number of ANN layers that fit on each chip. EfficientNetB4 is not pictured as it has over 60 convolutional layers (and several hundred other types of layers).}
    \label{fig:sparsity-layers}
\end{figure*}

\subsection{Latency Calculations}
\label{subsec:metric_calc}

We measure latency and throughput for single-input inference by calculating the total MAC or ACC operations per layer. To ensure consistency, we assume a 1-cycle latency for both operations. Each processing element (PE) computes in parallel, leveraging the parallelism inherent in neural network layers. Total on-chip latency per layer is derived using equations~(\ref{eq:ann_cycles_per_layer}) for ANN layers and~(\ref{eq:snn_cycles_per_layer}) for SNN layers, reflecting the computational workload distributed across PEs.

\begin{equation}
    \label{eq:ann_cycles_per_layer}
    cycles_{\rm ANN} = \frac{MACs \times cycles_{\rm MACs}}{ G \times \left \lceil{\frac{N}{G}}\right \rceil}  
\end{equation}

\begin{equation}
    \label{eq:snn_cycles_per_layer}
    cycles_{\rm SNN} = \frac{ACCs \times cycles_{\rm ACCs}}{ G \times \left \lceil{\frac{N}{G}}\right \rceil} ,
\end{equation}

\noindent where $N$ represents the number of neurons per layer and $G$ denotes the grouping of 256 neurons per core.

To incorporate the overhead of EMIO die-to-die latency, we account for a single packet transaction across the SerDes, estimated at 38 clock cycles for both input and output ports, resulting in a total of 76 cycles. The deserialization at the input port is pipelined, allowing the serial data stream to be expanded into parallel outputs during 38 of these 76 cycles. This pipelined deserialization is reflected in the calculation of latency overhead in (\ref{eq:EMIO_cycles_overhead}).

\begin{equation}
    \label{eq:EMIO_cycles_overhead}
    cycles_{\rm EMIO} = (\left \lfloor \frac{P_B}{N_c} \right \rfloor \times cycles_{\rm Ser}) + (P_B \times cycles_{\rm Des}),
\end{equation}

\noindent where $P_B$ denotes the packets transmitted across the chip boundary and $N_c$ represents the number of cores in the peripheral neural network layer.

The total latency is computed using (\ref{eq:total_cycles}), which sums the $L$-layer cycles of the SNN or ANN network and the $B$-layer EMIO cycles for layers crossing chip boundaries. This formulation accounts for both intra-chip computation and inter-chip communication overhead.

\begin{equation}
    \label{eq:total_cycles}
    cycles_{\rm Total} = \sum_{i=0}^{i=L} ( cycles_{\rm NN })  + \sum_{j=0}^{j=B} (cycles_{\rm EMIO})
\end{equation}

\subsection{Energy Consumption Calculations}
\label{subsec:intrachip_power}

We implemented an intra-chip energy model based on NN-Noxim and ORION 2.0 methodologies~\cite{NN-Noxim,orion2}, calculating energy consumption from the Tx and Rx transactions of routed and local packets, as defined in~\ref{eq:routed_packet_calc}. This modular framework estimates power usage for core components—including the PE, memory (MEM), and router—across ANN, SNN, and HNN NoC designs.

Energy costs were normalized relative to MAC operations on 45nm (8-bit precision) and 65nm (16-bit precision) CMOS technologies~\cite{dampfhoffer2022snns}. Within our 65nm power model, PE energy is reduced as SNN inference consumes approximately 0.06$\times$ the energy of a MAC operation, reflecting the efficiency of accumulations compared to multiplications. SRAM read/write costs are scaled accordingly, with ANN weights at 32-bit precision and SNN weights at 8-bit precision as defined in our architecture.

For die-to-die energy modeling, EMIO costs are calculated relative to intra-chip core-to-core routing. Based on findings from TrueNorth and ORION 2.0~\cite{merolla2014million, orion2}, die-to-die data movement consumes nearly 10$\times$ more energy than a MAC operation, and 224$\times$ that of a core-to-core packet per hop. We scale router link energy by this factor to estimate boundary packet costs in our models. As EMIO link energy is consistent for both ANN and SNN packets, interconnect energy depends solely on the packet volume, directly influencing latency and overall consumption. As models grow in size and depth, the combinatorial increase in the number of inter-die routing paths drives a superlinear increase in Router and EMIO energy, outpacing the near-linear growth of core compute and memory blocks.

\begin{figure*}[!h]
    \centering
    \begin{subfigure}[!h]{0.32\textwidth}
        \centering
        \includegraphics[width=\linewidth]{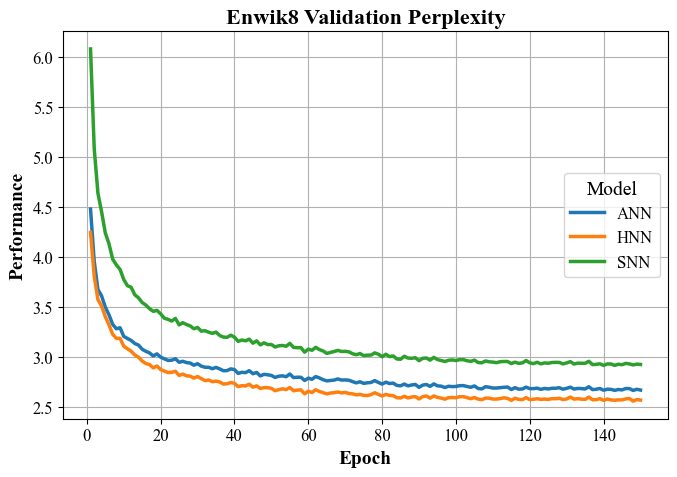}
        \caption{The validation perplexity in enwik8 dataset, lower is better. ANN, SNN, and HNN achieved a final perplexity of 2.66, 2.92, and 2.57, respectively.}
        \label{fig:enwik8}
    \end{subfigure}
    \hfill
    \begin{subfigure}[!h]{0.32\textwidth}
        \centering
        \includegraphics[width=\linewidth]{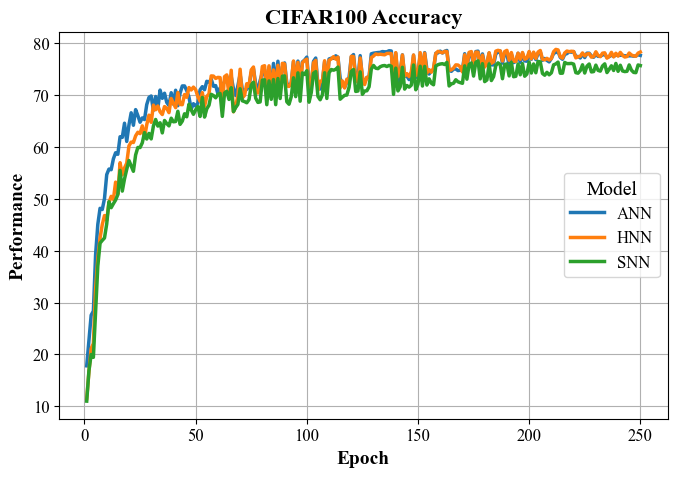}
        \caption{The testset accuracy in CIFAR100 dataset, higher is better. ANN, SNN, and HNN achieved a best classification accuracy of 78.65\%, 76.65\%, and 78.86\%, respectively.}
        \label{fig:cifar100}
    \end{subfigure}
    \hfill
    \begin{subfigure}[!h]{0.32\textwidth}
        \centering
        \includegraphics[width=\linewidth]{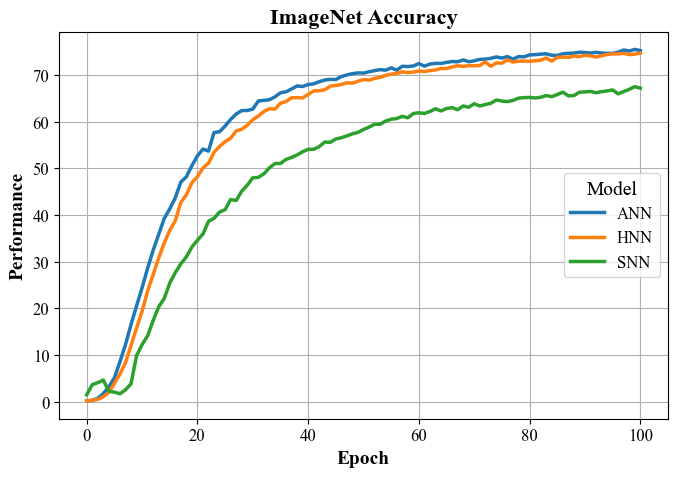}
        \caption{The test set accuracy in ImageNet-1K dataset, higher is better. ANN, SNN, and HNN achieved a best classification accuracy of 75.48\%, 67.50\%, and 74.78\% respectively.}
        \label{fig:imageNet}
    \end{subfigure}
    \caption{Comparison of performance across different datasets.}
    \label{fig:performance_metrics}
\end{figure*}

\section{Results}
\label{sec:results}

\subsection{Accuracy}
 
\begin{table} [!h]
    \centering
    \caption{Results on Enwik8, CIFAR100 and ImageNet-1K datasets. In Enwik8, we utilize character-level perplexity as the metric, lower is better. In CIFAR100 and ImageNet-1K, we employ top-1 classification accuracy as the metric, higher is better.}
    \label{tab:result}
    \begin{tabular}{c c c c}
        \hline
        & \textbf{ANN} & \textbf{SNN} & \textbf{HNN}  \\
        \hline
        \textbf{\# Enwik8 (PPL$\downarrow$)}
        & 2.66 & 2.92 & \textbf{2.57} \\
        \textbf{CIFAR100 (Top-1 Acc$\uparrow$)}
        & 78.65\% & 76.65\%  & \textbf{78.86\%} \\

        \textbf{ImageNet-1K (Top-1 Acc$\uparrow$)}
        & \textbf{75.48\%} & 67.50\% & 74.78\% \\
        \hline
    \end{tabular}
\end{table}

\begin{figure}[!h]
    \centering
    \includegraphics[width=1\linewidth]{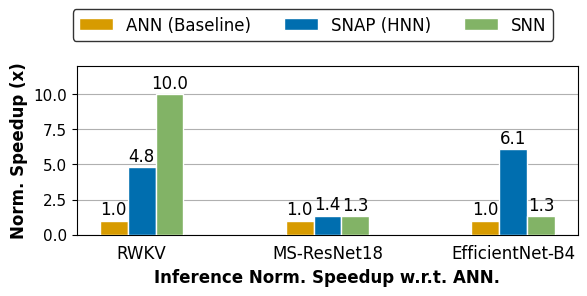}
    \caption{Latency per Inference Speedup ($\times$) for Enwik8, CIFAR100, \& ImageNet-1K inputs on corresponding RWKV, MS-ResNet18, \& EfficientNet-B4 model architectures using base parameters at 8-bit precision, 256 neuron grouping, \& 8 core chip NoC dimensions.}
    \label{fig:latency_speedup}
\end{figure}

\begin{figure*}[!h]
    \centering
        \includegraphics[width=\linewidth]{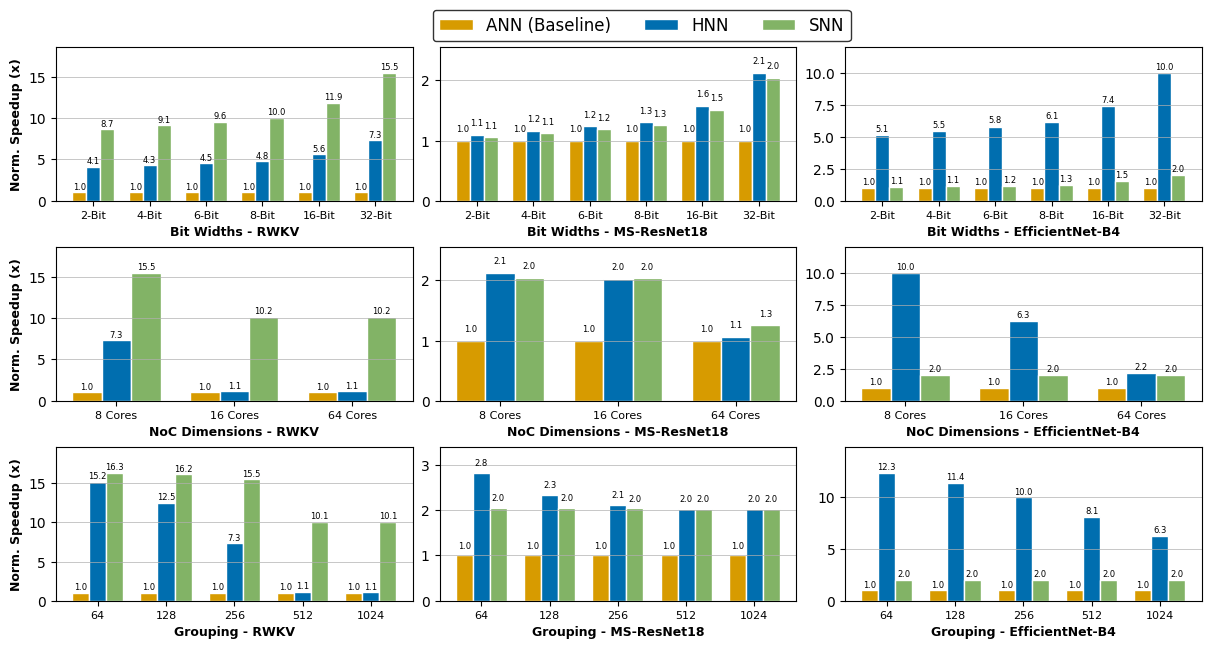}
        \caption{Normalized speed-up w.r.t. ANN as a function of bit-width, NoC dimensions, and grouping.}
        \label{fig:all_latency}
\end{figure*}

We employed MS-ResNet for image recognition and a six-layer, 512-size embedding RWKV model for language generation. For image recognition, experiments were conducted using the CIFAR100~\cite{krizhevsky2009learning} dataset with MS-ResNet18 and the ImageNet-1K~\cite{deng2009imagenet} dataset with EfficientNet-B4~\cite{tan2019efficientnet} modified to incorporate MS-ResNet blocks. For language generation, the Enwik8~\cite{Enwik8} dataset was used. Results are summarized in Tab.\ref{tab:result}. 

\textbf{Sparsity sweeping to find the Pareto-optimal point:} A sparsity sweep was applied to each model by applying a regularization term in the loss function that penalizes the model for excessive spiking:

\begin{equation}
    \mathcal{L} = \mathcal{L}_{CE} + \lambda\sum_i s_i,
\end{equation}

\noindent where $\mathcal{L}_{CE}$ is the cross-entropy loss, and the regularization term is weighted by $\lambda$, and only activated when the desired sparsity is exceeded in the training run. 
The results are shown in Fig.~\ref{fig:sparsity-sweep}, where HNN performance metrics (perplexity/accuracy) are highlighted in red and the latency improvement for all models is also displayed. Note that ANNs were not modulated for sparsity, and so their performance was consistent. 

\textbf{Sparsity breakdown per layer:} Average sparsity is a very limited metric, because it does not capture imbalanced sparsity, where high-firing layers can throttle overall performance. We provide a breakdown per layer in Fig.~\ref{fig:sparsity-layers}. Only the spiking layers in the HNN are accounted for, because zero-skipping is not implemented in the ANN cores. An interesting observation is that the distribution of spiking across the spike-based HNN layers is far more uniform than that of the SNNs, which means that HNNs are more stable and can reduce stalling between layers. 

The sparsity sweeps were used to identify the point of Pareto-optimality to be used for the full training run, and the final convergence curves are illustrated in Fig.~\ref{fig:performance_metrics}.

Our findings show that HNNs outperform SNNs in both image classification and language generation tasks. Remarkably, HNNs also marginally outperform ANNs on CIFAR100 and Enwik8, while maintaining competitive performance on ImageNet-1K, outperforming SNNs by 7.28\%. The superior performance of HNNs over SNNs is expected; however, the observed outperformance of HNNs over ANNs on CIFAR100 and Enwik8 is noteworthy. We hypothesize that SNN components in HNNs may act as a form of regularization, reducing overfitting on these datasets. For instance, on CIFAR100, ANNs demonstrated faster convergence compared to SNNs and HNNs, but HNNs ultimately achieved comparable or better performance. Similarly, on the larger ImageNet-1K dataset, which mitigates randomization effects due to its scale, HNNs maintained competitive results. These results highlight the potential of HNNs to surpass both SNNs and ANNs, particularly in tasks where overfitting is a concern, suggesting their viability for larger-scale applications.

\begin{figure}[!htbp]
    \centering
    \includegraphics[width=0.8\linewidth]{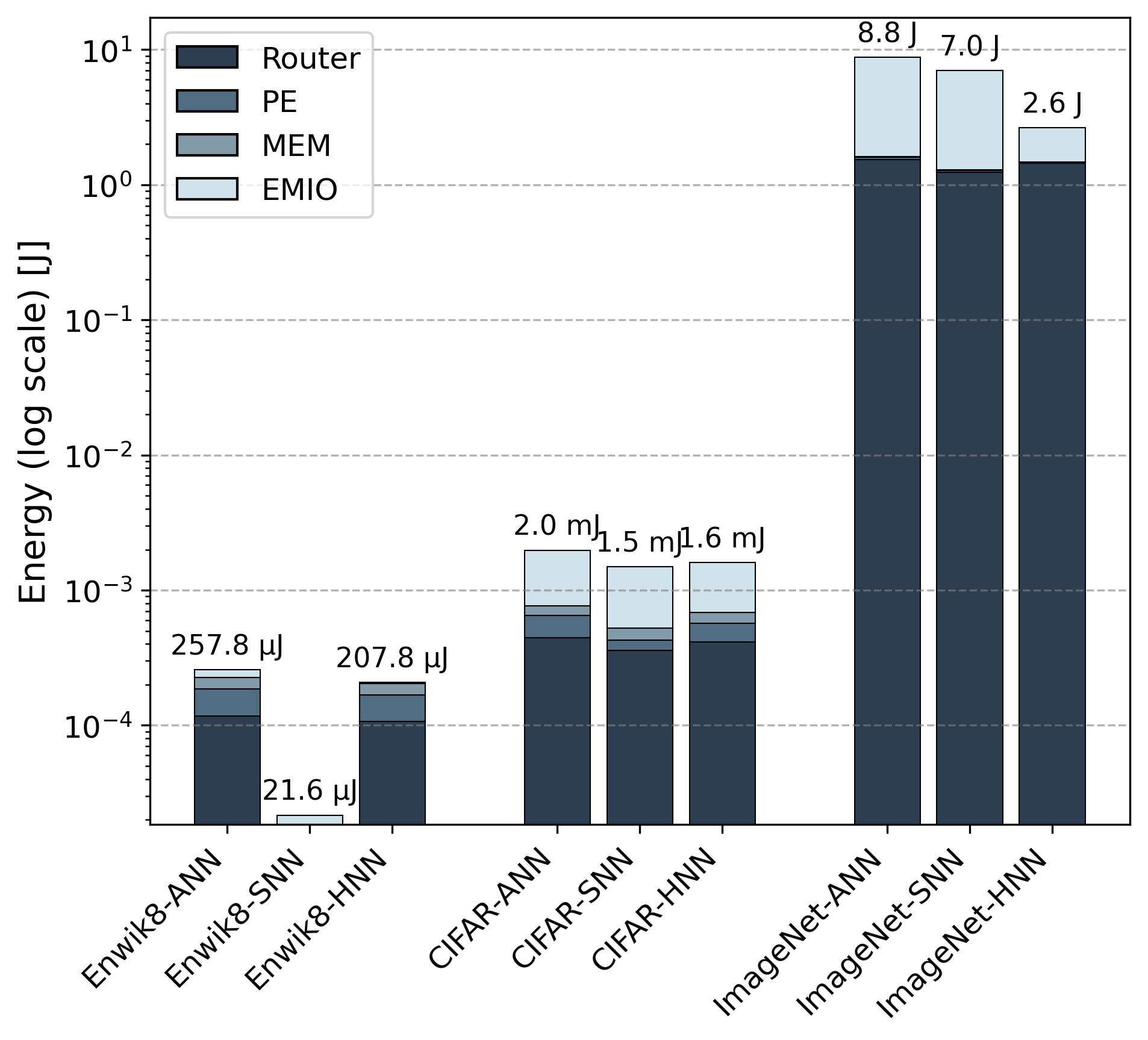}
    \caption{Energy Consumption (J) per Inference for Enwik8, CIFAR100, \& ImageNet-1K on corresponding RWKV, MS-ResNet18, \& EfficientNet-B4 models.}
    \label{fig:energy}
\end{figure}

\begin{figure*}[!h]
    \centering
    \includegraphics[width=\linewidth]{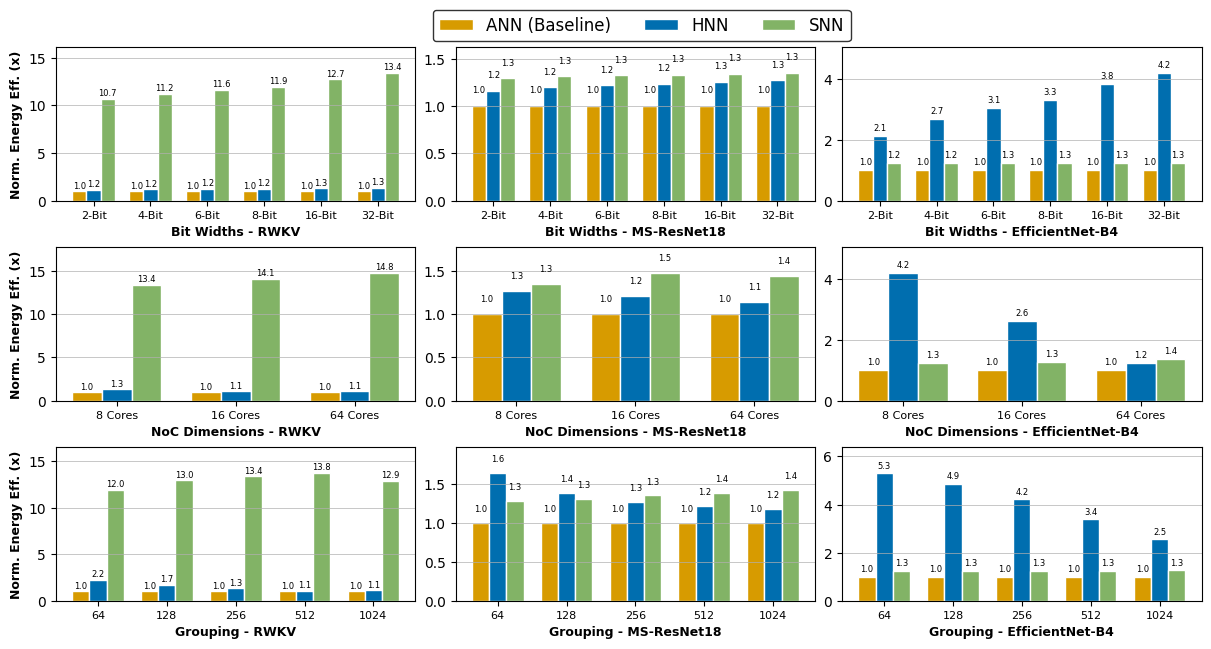}
    \caption{Normalized energy efficiency w.r.t ANN as a function bit-width, NoC dimensions, and grouping.}
    \label{fig:all_energy}
\end{figure*}

\subsection{Latency}
\label{subsec:Latency}

We mapped all software models to approximate workloads on the proposed ANN, SNN, and HNN designs, detailing the neuron, core, and chip resources utilized by each model. Using 8-bit precision, 256-neuron grouping, and an 8-core NoC configuration as a baseline, Fig.~\ref{fig:latency_speedup} presents our latency and accuracy reference metrics.

Across varying bit precisions, neuron groupings, and NoC dimensions, results demonstrate that HNNs achieve the fastest inference latency on static datasets, while SNNs maintain an advantage on dynamic datasets due to their reduced timesteps for achieving high accuracy. As shown in Fig.~\ref{fig:all_energy}, the HNN exhibited speedups ranging from 1.1$\times$ to 15.2$\times$ over ANN models on benchmark datasets. Notably, as model resources scaled or bit-precision increased and die-to-die communication demands increased, HNNs demonstrated faster inference latency than both ANNs and SNNs, highlighting their scalability and efficiency in large-scale applications.

\subsection{Energy}

We found that the HNN baseline model was 1$\times$ to 3.3$\times$ more energy efficient than the ANN design per inference. An energy breakdown per component is shown in Fig.~\ref{fig:energy}. This efficiency arises from the computational cost reduction inherent in SNN layers, which allow HNNs to lower the total energy consumption per operation. As model sizes increase, the performance margin of the HNN further improves. While sweeping across other parameter configurations, energy efficiency gains continue up to 5.3$\times$ using a smaller neuron-to-processing-element grouping, which corresponds to allocating more resources per network. 

For example, deploying EfficientNet-B4 on the ImageNet-1K dataset required 329 times more chips than RWKV and approximately 73 times more chips than MS-ResNet-18, significantly increasing the number of die-to-die channels utilizing spike-based conversions. This scalability makes HNNs increasingly effective for large-scale models, avoiding the accuracy degradation commonly observed in purely SNN designs. This also explains why the HNN has the lowest margin of improvement for the RWKV 6-layer model, but shows promising scaling. The HNN model also reduced router energy consumption in static data in comparison to the SNN model by limiting spiked-based communication to only peripheral traffic which requires a spike train of $T$ spikes per activation sent.

\section{Conclusion}
\label{sec:conclusion}

As deep learning workloads continue to scale, computational demands increasingly exceed the rate at which data can be transferred on-chip, creating significant bottlenecks. SNNs, with their sparse, spike-based activations, offer reduced communication traffic and improved energy efficiency compared to conventional ANNs.

In this paper, we introduced an approach to HNN-based co-design that mitigates these bottlenecks by combining the strengths of SNNs and ANNs. By employing sparse SNN neurons at the periphery of the chip, near I/O bottlenecks, and dense ANN neurons within the chip interior, we balance the trade-offs between accuracy and communication efficiency. This partitioned design allows the ANN layers to deliver high accuracy, while the SNN layers significantly alleviate bandwidth constraints across chip boundaries and within the NoC.

Our evaluations demonstrate that the HNN can achieve up to 15.2$\times$ speedup over traditional ANN architectures on static data, using a smaller neuron-to-processing-element grouping, while maintaining similar accuracy and significantly improving energy efficiency. The improvements continue to scale relative to ANN-only accelerators, given the larger number of chip boundaries or higher bit-precision data that induce packet congestion. These results highlight the potential of hybrid architectures to address the scalability challenges of modern deep learning workloads, particularly in bandwidth-constrained environments.

Future work will focus on extending the HNN’s capabilities to larger-scale systems and exploring its applicability to emerging workloads, including transformers and graph neural networks, further establishing its role as a scalable and energy-efficient solution for advanced deep learning architectures.

\section*{Acknowledgments}
Redacted for double blind peer review.


\bibliographystyle{ACM-Reference-Format}
\bibliography{refs}


\end{document}